\documentclass{article}
\usepackage[top=1in, bottom=1in, left=1in, right=1in]{geometry}
\usepackage{planargraph}
\usepackage{graphicx}
\usepackage{amsmath,amssymb,amsthm}
\usepackage{theoremsetc}
\usepackage{algorithm}
\usepackage[noend]{algorithmic}
\usepackage{enumitem}

\newcommand{\DIST}{\textrm{D}}

\newcommand{\inflow}{\textrm{inflow}}
\newcommand{\outflow}{\textrm{outflow}}
\newcommand{\proc}[1]{\textnormal{\scshape#1}}

\newcommand{\sources}{\emph{sources lemma}}
\newcommand{\sinks}{\emph{sinks lemma}}

\begin{document}
\begin{titlepage}

\title{Multiple-Source Multiple-Sink Maximum Flow in Directed Planar Graphs in Near-Linear Time}

\author{Glencora Borradaile \\ Oregon State University \and 
Philip N. Klein \\Brown University \and
Shay Mozes\\Brown University \and
Yahav Nussbaum \\Tel-Aviv University \and
Christian Wulff-Nilsen \\ Carleton University}

\maketitle
\thispagestyle{empty}
\begin{abstract}
  We give an $O(n \log^3 n)$ algorithm that, given an $n$-node
  directed planar graph with arc capacities, a set of source nodes, and
  a set of sink nodes, finds a maximum flow from the sources to the
  sinks.  
Previously, the fastest algorithms known for this problem were those for general graphs.
\end{abstract}

\end{titlepage}
\section{Introduction}
The maximum flow problem
with multiple sources and sinks in a directed graph with arc-capacities is, informally, to find a
way to route a single commodity from a given set of
sources to a given set of sinks such that the total amount of the
commodity that is delivered to the sinks is maximum subject to each
arc carrying no more than its capacity. In this paper we
study this problem in planar graphs.

The study of maximum flow in planar graphs has a long history.  In
1956, Ford and Fulkerson introduced the max $st$-flow problem, gave a
generic augmenting-path algorithm, and also gave a particular
augmenting-path algorithm for the case of a planar graph where $s$ and
$t$ are on the same face 
Researchers have since published many
algorithmic results proving running-time bounds on max $st$-flow for
(a) planar graphs where $s$ and $t$ are on the same face, (b)
undirected planar graphs where $s$ and $t$ are arbitrary, and (c)
directed planar graphs where $s$ and $t$ are arbitrary.  The best
bounds known are (a) $O(n)$~\cite{HKRS97}, (b) $O(n \log \log
n)$~\cite{INSW11}, and (c) $O(n \log n)$~\cite{BorradaileK09},
where $n$ is the number of nodes in the graph.

Maximum flow in planar graphs  with multiple sources and sinks was
studied by Miller and Naor \cite{MN95}. When it is known 
how much of the commodity is produced/consumed at each source and each
sink, finding a consistent routing of flow that respects arc
capacities can be reduced to \emph{negative-length shortest paths},
which we now know can be
solved in planar graphs in $O(n \log ^2 n / \log \log n)$ time
\cite{MWN10}. Otherwise,  Miller and
Naor gave an $O(n \log^{3/2} n)$ 
algorithm for the case where all the sinks and the sources are on the
boundary of a single face, and generalized it to an $O(k^2 n^{3/2}
\log^2 n)$-time algorithm for the case where the sources and the sinks
reside on the boundaries of $k$ different faces.\footnote{The time bound of the
first algorithm can be improved to $O(n \log n)$ using the linear-time
shortest-path algorithm of Henzinger et al.~\cite{HKRS97}, and the
time bound of the second algorithm can be improved to $O(k^2 n \log^2
n)$ using the $O(n \log n)$-time single-source single-sink maximum
flow algorithm of Borradaile and Klein \cite{BorradaileK09}.}

However, the problem of maximum flow with multiple sources and sinks
in planar graphs without any additional restrictions remained open.
In general (i.e., non-planar) graphs, multiple sources and sinks
can be reduced to the single-source single-sink case by introducing an
artificial source and sink and connecting them to all the sources and
sinks, respectively---but this reduction does not preserve planarity. For more
than twenty years since the problem was explicitly stated and
considered~\cite{MN95}, the fastest known algorithm for computing
multiple-source multiple-sink max-flow in a planar graph has been to use
this reduction in conjunction with a general maximum-flow algorithm
such as that of Sleator and Tarjan~\cite{ST83} which leads to a
running time of $O(n^2 \log n)$.  For integer capacities less than
$U$, one could instead use the algorithm of Goldberg and
Rao~\cite{GR98}, which leads to a running time of $O(n^{1.5} \log n
\log U)$. No planarity-exploiting algorithm was known for the problem.

\paragraph{Our Result}
The main result of this paper is an algorithm for the problem that is
optimal up to a small poly-logarithmic factor.

\begin{theorem}\label{main-thm}
There exists an algorithm that solves the maximum flow problem with
multiple sources and sinks in an $n$-node directed planar graph in $O(n \log^3 n)$ time.
\end{theorem}

\paragraph{Application to computer vision problems}
Multiple-source multiple-sink min-cut arises in addressing a family of
problems associated with the terms {\em metric labeling}
(Kleinberg and Tardos, \cite{KleinbergT99}), {\em Markov Random Fields} \cite{GemanGeman}, and
{\em Potts Model} (see also~\cite{BVZ01,Hochbaum01}).  In low-level
vision problems such as {\em image
restoration}, {\em segmentation}, {\em stereo}, and {\em motion}, the goal is to assign labels from a
set to pixels so as to minimize a penalty function.  The penalty
function is a sum of two parts.  One part, the {\em data component},
has a term for each pixel; the cost depends on the discrepancy between 
the observed data for the pixel and the label chosen for it.  The
other part, the {\em smoothing component}, penalizes neighboring
pixels that are assigned different labels.

For the {\em binary} case (when the set of available labels has size
two), finding the optimal solution is reducible to multiple-source
multiple-sink min-cut.
\cite{Greig}.  For the case of more than two labels, there is a
powerful and effective heuristic \cite{BVZ01} using
very-large-neighborhood \cite{AEOP02} local search; the inner loop
consists of solving the two-label case.  The running time for solving
the two-label case is therefore quite important.  For this reason,
researchers in computer vision have proposed new algorithms for max
flow and done experimental studies comparing the run-times of
different max-flow algorithms on the instances arising in this
context~\cite{BK04,KohliT07}.  All of this is evidence for the
importance of the problem.

For the (common) case where the underlying graph of pixels is the
two-dimensional grid graph, 
our result yields a theoretical speed-up for this important
computer-vision subroutine.\footnote{Note that the single-source,
  single-sink max-flow algorithm of~\cite{BorradaileK09} was
  implemented by computer-vision researchers~\cite{Schmidt09} and found to
  be useful in computer vision and to be faster than the competitors.}

Hochbaum~\cite{Hochbaum01} describes a special case of the penalty function in which the
data component is convex and the smoothing component is linear; in
this case, she shows that an optimal solution can be found in time
$O(T(m, n)+n \log U)$ where $U$ is the maximum label, and $T(m, n)$ is the
time for finding a minimum cut.  She mentions specifically 
image segmentation, for which the graph is planar.  For this case, by using our
algorithm, the optimal solution can be found in nearly linear time

\paragraph{Application to maximum bipartite matching}
Consider the problem of maximum matching in a bipartite planar
graph.  It is well-known how to reduce this problem to
multiple-source, multiple-sink maximum flow.  Our result is the
first planarity-exploiting algorithm for this problem (and the first
near-linear one).

\paragraph{Techniques}

To obtain our result, we employ a wide range of sophisticated
algorithmic techniques for planar graphs, 
some of which we adapted to our needs
while others are used unchanged.
Our algorithm uses pseudoflows~\cite{GoldbergT87,Hochbaum08} and a
divide-and-conquer scheme influenced by that of~\cite{JV82} and
that of~\cite{MN95}.  
We adapt a method for using shortest paths to solve max $st$-flow when $s$ and $t$ are
adjacent~\cite{Hassin81}, and a data structure for implementing Dijkstra in
a dense distance graph derived from a planar graph~\cite{FR06}. Among
the other techniques we employ are: using cycle separators~\cite{Miller86} recursively while
keeping the boundary nodes on a constant number of
faces~\cite{KleSub98,Subramanian95,FR06}, an algorithm for
single-source single-sink max flow~\cite{BorradaileK09,Erickson10}, 
an algorithm for computing multiple-source shortest
paths~\cite{Klein05,CabelloC07}, a method for cancelling cycles of
flow in a planar graph~\cite{KaplanNussbaum2009}, an algorithm for
finding shortest paths in planar directed graphs with negative lengths~\cite{KleinMW10,MWN10},
and a data structure for range queries in a Monge matrix~\cite{KS11}.

\subsection{Preliminaries}\label{sec:prel}

We assume the reader is familiar with the basic definitions of planar embedded graphs and their duals (cf. ~\cite{BorradaileK09}).
Let $G=(V,E)$ be a planar embedded graph with node-set $V$ and arc-set $E$. 
We use the term \emph{arc} to emphasize that edges are directed. 
The term \emph{edge} is used when the direction of an arc is not important. 
For each arc $a$ in the arc-set $E$, 
we define two oppositely directed {\em darts}, one in the same orientation as $a$ (which 
we sometimes identify with $a$) and one in the opposite orientation~\cite{BorradaileK09}. 
We define $\rev(\cdot)$ to be the function that takes each dart to the corresponding dart in 
the opposite direction.
It is notationally convenient to equate the edges, arcs and darts of $G$ with the edges, arcs and darts of the dual $G^*$.

Let $S \subset V$ be a set of nodes called sources, and let $T \subseteq V \setminus S$ be a set of nodes called sinks.

A {\em flow assignment} $\fvec(\cdot)$ is a real-valued function
on darts satisfying {\em antisymmetry}:
\begin{equation*} \label{eq:antisymmetry}
\fvec(\rev(d))= -\fvec(d)
\end{equation*}

A {\em capacity assignment} $\cvec(\cdot)$ is a real-valued function on darts. 

A flow assignment $\fvec(\cdot)$ \emph{respects the capacity} of dart $d$
if $\fvec(d) \leq \cvec(d)$. $\fvec(\cdot)$ is called a \emph{pseudoflow} if it
respects the capacities of all darts. 

For a given flow assignment $\fvec(\cdot)$, the {\em net inflow} (or just
{\em inflow}) node $v$ is
$\inflow_{\fvec}(v) = \sum_{\textrm{dart } d: \head(d) = v} \fvec(d)$\footnote{An equivalent definition, in terms of arcs, is 
$\inflow_{\fvec}(v) = \sum_{a \in E : \head(a) =v} \fvec(a) - \sum_{a \in E :
  \tail(a) = v} \fvec(a) $.}.
The {\em outflow} of $v$ is
$\outflow_{\fvec}(v) = - \inflow_{\fvec}(v)$.
A flow assignment $\fvec(\cdot)$ is said to \emph{obey conservation} at
node $v$ if $\inflow_{\fvec}(v) = 0$.
A \emph{feasible flow} is a pseudoflow that obeys conservation at every node other than the sources and sinks. 
A \emph{feasible circulation}  is a pseudoflow that obeys conservation at all nodes.
The {\em value} of a feasible flow $\fvec(\cdot)$ is the
 sum of inflow at the sinks, $\sum_{t\in T}\inflow_{\fvec}(t)$ or, equivalently, the sum of outflow at the sources.
The maximum flow problem is that of finding a feasible flow with maximum value.

For two flow assignments $\fvec,\fvec'$, the \emph{addition} $\fvec+\fvec'$ is the flow that
assigns $\fvec(d)+\fvec'(d)$ to every dart $d$.

A \emph{residual path} in $G$ is a path whose darts all have strictly positive capacities.
For two sets of nodes $A,B$, $A \stackrel{G}{\rightarrow} B$ is used
to denote the existence of some residual $a$-to-$b$ path in $G$ for 
some nodes $a \in A$ and $b \in B$. Conversely, $A
\stackrel{G}{\nrightarrow} B$ is used to denote that no such path
exists. We will omit the graph $G$ when it is clear from the context.

The \emph{residual graph} of $G$ with respect to a flow assignment $\fvec(\cdot)$ is the graph $G_{\fvec}$ with
the same arc-set, node-set, sources and sinks, and with capacity assignment $\cvec_{\fvec}(\cdot)$
such that for every dart $d$, $\cvec_{\fvec}(d) = \cvec(d) - {\fvec}(d)$.
It is well known that a feasible flow $\fvec$ in $G$ is maximum if and only if $S \stackrel{G_{{\fvec}}}{\nrightarrow} T$.

Let $\fvec$ be a pseudoflow in a planar graph $G$. Let
$V^+$ denote the set of nodes $\{v \in V \setminus (S \cup T) :
\inflow_{\fvec}(v) > 0\}$. Similarly, let
$V^-$ denote the set of nodes $\{v \in V \setminus (S \cup T) :
\inflow_{\fvec} (v) < 0\}$.
Suppose $S \cup V^+ \stackrel{G_{\fvec}}\nrightarrow T \cup V^-$.
For a graph with $n$ nodes and $m$ edges, 
there exists an $O(m \log n)$-time algorithm that converts the pseudoflow $\fvec$ into
a maximum feasible flow $\fvec'$~\cite{JV82, ST83}. In planar graphs, this can be done in linear time
by first canceling flow 
cycles using the technique of Kaplan and
Nussbaum~\cite{KaplanNussbaum2009}, and then
by sending back flow from $V^+$ and into $V^-$ in topological sort
order. See Appendix~\ref{apn:psuedoflow-to-flow} for details.


\subsection{Overview of the algorithm} \label{sec:overview}

Consider the following recursive approach for finding a maximum
multiple-source multiple-sink flow: split the input graph $G$
in two using a simple cycle separator $C$~\cite{Miller86} and recursively solve the
max flow problem in the two subgraphs. When the two recursive calls
have been executed, in each of the two subgraphs there is no residual
path from any source to any sink. If we further make sure that in each
of the two subgraphs there is no residual path from any source to $C$
and from $C$ to any sink, then, since $C$ is a separator, there is no
residual path from any source to any sink in $G$.

We therefore solve a slightly more general problem recursively in the
two subgraphs: roughly speaking, find a flow such that there is no
residual path from a source to a sink or to $C$ and no residual path
from $C$ to a sink (Section~\ref{sec:algo}).  After the two
recursive calls there is no residual path from any source to any sink
in $G$. However, the requirement that there is no residual path from
any source to $C$ and from $C$ to any sink cannot be achieved by a
feasible flow but rather by a pseudoflow in which there might be 
excess inflow or excess outflow on nodes of $C$. We deal with this by
solving a new max flow problem where nodes of $C$ are treated as
sources and sinks, limited in supply/demand by their excess
inflow/outflow (Section~\ref{sec:fix}).

We exploit a relation between primal circulations and dual shortest
paths to maintain a succinct representation of the flow during
critical phases of the algorithm, using the fact that there are only
$O(\sqrt n)$ sources and sinks, all cyclically ordered on $C$ (Section
~\ref{sec:eff}). 
Even though
our representation does not explicitly store the flow on nearly any
arc in the graph, we can augment it efficiently towards optimality while
maintaining feasibility. An important tool we use is Fakcharoenphol
and Rao's efficient implementation of Dijkstra's
algorithm~\cite{FR06}, which we adapt to our needs.

The
resulting pseudoflow can then be turned into a max flow in linear time
using existing techniques (Appendix~\ref{apn:psuedoflow-to-flow}). This leads to an $O(n\log^3n)$
time algorithm for max flow.

\section{The Algorithm}\label{sec:algo}

Now we describe the algorithm, referred to as \proc{MultipleSourceMultipleSinkMaxFlow}, in more detail.
In order to treat nodes of the cycle separator both as sources and as sinks in recursive
calls, we introduce a new node set $A$. At the top recursion level, $A
= \emptyset$. 
In general, $A$ has constant size;
more precisely $|A|\leq 6$.
\begin{figure*}[h!!]
\proc{MultipleSourceMultipleSinkMaxFlow}($G,\cvec$,$S$,$T$,$A$)\\
{\bf Input:} a directed planar graph $G$ with non-negative
capacities $\cvec$, a set $S$ of source nodes, a set $T$ of sink
nodes, a set $A$ of at most six nodes \\ 
{\bf Output:} a pseudoflow $\fvec$ obeying conservation
everywhere but $S,T,A$ and s.t.~$S \stackrel{G_{\fvec}} \nrightarrow T$, 
$S \stackrel{G_{\fvec}} \nrightarrow A$, 
$A \stackrel{G_{\fvec}} \nrightarrow T$.
\end{figure*}

The algorithm finds a simple cycle separator $C$~\cite{Miller86} and contracts all
edges of $C$ except one. This, essentially, merges all the nodes of
$C$ into a single supernode $v$, and turns $C$ into a self loop.
For simplicity of presentation we assume that no sources or sinks lie
on $C$.\footnote{This does not lose generality since if a node $u \in C$ is
a source, one can introduce a new node $u'$ and an arc $u'u$ whose capacity
equals the sum of capacities of arcs outgoing from $u$. Now consider
$u'$ as a source instead of $u$. Since the separator has just $O(\sqrt
n)$ nodes, this will not affect the running time. Sinks can be handled in a similar fashion.}
The algorithm then recursively solves the problem 
on the subgraphs enclosed and not enclosed by that self loop (the self
loop itself need not be included in any of the subgraphs), adding the
supernode $v$ that represents $C$ 
to the set $A$. 
In order to keep the cardinality of $A$
at most six, the algorithm alternately applies the cycle separator theorem with weights uniformly distributed on
all nodes and uniformly distributed on only the nodes of $A$. This
technique is similar to that used in~\cite{KleSub98,Subramanian95,FR06}.

After the recursive calls, the algorithm 
uncontracts the edges of $C$.
At this stage there are no residual paths between sources and sinks in
the entire graph, but there might be 
excess inflow (positive or negative) at the nodes of $C$. 
The algorithm then calls the procedure
\proc{FixConservationOnPath} that pushes flow between the nodes of $C$ so
that there are no residual paths between nodes of $C$  with positive
inflow and nodes of $C$ with negative inflow 
(the path in the name of the procedure is the cycle $C$ without one edge). 
This procedure is discussed in Section~\ref{sec:fix}; the interface is:
\begin{figure*}[h!!]{\proc{FixConservationOnPath}$(G,P, \cvec, \fvec_0)$}\\
{\bf Input:} a directed planar graph $G$, simple path
$P$, capacity function $\cvec$, and a pseudoflow $\fvec_0$\\
{\bf Output:} a pseudoflow $\fvec$ s.t. 
{\it (i)} $\fvec-\fvec_0$ satisfies conservation everywhere but $P$, and \\
{\it (ii)} $\{ v \in P :
\inflow(v)>0\}\stackrel{G_{\fvec}}{\nrightarrow} \{ v \in P : \inflow(v)<0\}$.\\
{\bf Running Time:} $O(n \log^2 n / \log\log n + |P|^2 \log^2 n)$
\end{figure*}

\vspace{-5pt}
Next, the algorithm iterates over the nodes $a_i$ of $A$. The
algorithm calls the procedure \proc{CycleToSingleSinkLimitedMaxFlow}
that, roughly speaking, pushes as much excess flow as possible from
$C$ to $a_i$.  If $C_i^+$ is the set of nodes of $C$ that are
reachable via residual paths from some node of $C$ with positive
inflow at the beginning of iteration $i$,
\proc{CycleToSingleSinkLimitedMaxFlow} pushes flow among the nodes of
$C_i^+$ and from the nodes of $C_i^+$ to $a_i$.  The result is that
remaining inflow at nodes of $C_i^+$ is non-negative and there
are no residual paths from nodes of $C$ with positive inflow to
$a_i$. See Section~\ref{sec:C2A}; the
interface is:
\begin{figure*}[h!]
\proc{CycleToSingleSinkLimitedMaxFlow}($G,\cvec, \fvec_0, C, a_i$) \\
{\bf Input:} a directed planar graph $G$ with capacities $\cvec$, a pseudoflow $\fvec_0$, a simple cycle $C$, a sink $a_i$. \\
{\bf Assumes:}  $\forall v \in C^+,\ \inflow_{\fvec_0}(v) \geq 0$, 
 where 
$C^+=\{ v \in C : \{x\in C : \inflow_{\fvec_0} (x)>0\} \stackrel{G_{\fvec_0}}{\rightarrow} v\}$.\\
{\bf Output:} a pseudoflow $\fvec$ s.t.
{\it(i)} $\fvec - \fvec_0$ obeys conservation everywhere but
$C^+ \cup \{t\}$, 
{\it(ii)} $\forall v \in C^+,\ \inflow_{\fvec}(v) \geq 0$, 
{\it (iii)} $\{ v \in C : \inflow_{\fvec}(v) > 0\}
\stackrel{G_{\fvec}}{\nrightarrow} a_i$.\\
{\bf Running Time:} $O(n \log^2 n / \log\log n + |C|^2 \log^2 n)$.
\end{figure*}

\vspace{-5pt} A similar procedure \proc{SingleSourceToCycleLimitedMaxFlow} is called
to push flow from $a_i$ to $C$ to eliminate as much negative
inflow as possible (using a similarly defined set $C_i^-$).

Finally, the algorithm pushes back flow from any nodes of $C$ with
positive inflow to $S$ and pushes flow back from $T$ into any nodes of $C$ with
negative inflow.

\begin{algorithm}[h!!]\caption{\proc{MultipleSourceMultipleSinkMaxFlow}($G,\cvec$,$S$,$T$,$A$) \label{alg:msms}}
{\bf Input:} a directed planar graph $G$ with non-negative
capacities $\cvec$, a set $S$ of source nodes, a set $T$ of sink nodes, a set $A$ consisting of a constant number of nodes $A=\{a_i\}_{i=1}^k$. \\
{\bf Output:} a pseudoflow $\fvec$ obeying conservation
everywhere except $S,T,A$ and s.t.~$S \stackrel{G_{\fvec}} \nrightarrow T$, 
$S \stackrel{G_{\fvec}} \nrightarrow A$, 
$A \stackrel{G_{\fvec}} \nrightarrow T$.\\
\begin{algorithmic}[1]
\vspace{-10pt}
\STATE add zero capacity arcs to triangulate and 2-connect $G$
(required for simple cycle separators)
\STATE find a balanced (w.r.t.~$|G|$ and $|A|$, alternately) cycle separator $C$ in $G$  disjoint from $S$ and $T$
\STATE let $P$ be a path comprising of all of $C$'s edges except one
edge $e$ \label{line:P}
\STATE contract all the edges of $P$, turning $e$ into a self loop
incident to the only remaining node $v$ of $C$ \label{line:contract}
\STATE let $G_1$ and $G_2$ be the subgraph of $G$ enclosed  and not
enclosed by $e$, respectively
 \STATE $\fvec := $ \proc{MultipleSourceMultipleSinkMaxFlow}($G_1, \cvec_{|G_1}, S\cap G_1, T\cap G_1, (A \cap G_1) \cup \{v\}$) \label{line:recurse1} 
 \STATE $\fvec := \fvec \ + $ \proc{MultipleSourceMultipleSinkMaxFlow}($G_2, \cvec_{|G_2}, S\cap G_2, T\cap G_2, (A \cap G_2) \cup \{v\}$) \label{line:recurse} 
\STATE uncontract the edges of $P$\label{line:uncontract}
\STATE $\fvec :=$ \proc{FixConservationOnPath}($G,P,\cvec,\fvec$) \label{line:fix} 
\FOR {$i=1,2, \dots, k$}\label{line:i-loop}
 \STATE $\fvec :=$ \proc{CycleToSingleSinkLimitedMaxFlow}($G,\cvec, \fvec, C, a_i$)\label{line:to-ai}
 \STATE $\fvec :=$ \proc{SingleSourceToCycleLimitedMaxFlow}($G,\cvec, \fvec, a_i, C$)\label{line:from-ai}
\ENDFOR
\STATE push positive excess from $C$ to $S$ and negative excess to $C$
from $T$\label{line:pseudo}
\RETURN $\fvec$
\end{algorithmic}
\end{algorithm}

\paragraph{Correctness of Algorithm~\ref{alg:msms}}
The correctness of Algorithm~\ref{alg:msms} is proved in detail in Appendix~\ref{sec:correct}. 
The proof consists of a sequence of lemmas that  prove
that each step of the algorithm eliminates some undesired residual paths
without reintroducing undesired residual paths. The arguments used are elementary.

\paragraph{Running Time Analysis}
The number of nodes of the separator cycle $C$ used to partition $G$
into $G_1$ and $G_2$ is $O(\sqrt{|G|})$. Therefore, each
invocation of \proc{FixConservationOnPath},
\proc{CycleToSingleSinkLimitedMaxFlow} and \proc{SingleSourceToCycleLimitedMaxFlow} in $G$
takes $O(|G|\log^2 |G| / \log\log |G| + |C|^2 \log^2 |C|) = O(|G| \log^2|G|)$ time.

The way we recursively partition into subgraphs is very similar to that of Fakcharoenphol and Rao~\cite{FR06}. 
In their
algorithm, they spend $O(|G'|\log^2|G'|)$ time on each subgraph $G'$
in the recursive decomposition and prove a total time bound
of $O(n\log^3n)$. 
By the same arguments, our algorithm runs in $O(n\log^3n)$ time.

\section{Eliminating Residual Paths Between Nodes with Positive Inflow
  and Nodes with Negative Inflow on a Path} \label{sec:fix}

In this section we present an efficient
implementation of the fixing procedure which, roughly speaking, given
a path with nodes having positive, negative, or zero inflow,
pushes flow between the nodes of the path so
that eventually there are no
residual paths from nodes with positive inflow to nodes with negative inflow.

We begin by describing an abstract algorithm for the fixing procedure.
The abstract algorithm is given as Algorithm~\ref{alg:fix}.
It is similar to a technique used by Venkatesan and Johnson~\cite{JV82}.
Let $M$ be the sum of capacities of all of the darts of $G$. 
The algorithm first increases the capacities of darts of the path $P$ and their
reverses by $M$. Let $p_1, p_2, \dots, p_{k+1}$ be the nodes of $P$.
The algorithm processes the nodes of $P$ one after the other. 
Processing $p_i$ consists of decreasing the capacities of $d_i=p_ip_{i+1}$ and
$rev(d_i)$ by $M$ (i.e., back to their original capacities), and
trying to eliminate positive inflow $x$ at $p_i$ by pushing at most $x$
units of flow from $p_i$ to $p_{i+1}$.
The intuition for doing so is that the flow after the push either
obeys conservation at $p_i$ or there are no residual paths from $p_i$
to any of the other nodes of $P$ (this is where we use the large
capacities on the darts between unprocessed nodes). See appendix~\ref{sec:abs-cor}
for a formal proof of correctness.
Negative inflow at $p_i$ is handled in a similar manner by pushing
flow from $p_{i+1}$ to $p_i$.

\begin{algorithm}[h!]\caption{\proc{AbstractFixConservationOnPath}$(G,P, \cvec, \fvec_0)$} \label{alg:fix}
{\bf Input:} directed planar graph $G$, simple path
$P=d_1 d_2\ldots d_k$, capacity function $\cvec$, and pseudoflow $\fvec_0$\\
{\bf Output:} a pseudoflow $\fvec'$ s.t.
{\it(i)} $\fvec'-\fvec_0$ satisfies conservation at nodes not on $P$, and 
{\it (ii)} with respect to $\fvec'$, there are no residual paths from nodes
  of $P$ with positive inflow to nodes of $P$ with negative inflow. \\
\begin{algorithmic}[1]
\STATE \vspace{-10pt} $\fvec' = \fvec_0$\label{0-1}
\STATE $\cvec[d]  = \cvec[d] + M$ for all darts $d$ of $P \cup \rev(P)$ \vspace{5pt}
\FOR {$i=1,2, \ldots, k$ } \label{0-loop} 
   \STATE let $p_i$ and $p_{i+1}$ be the tail and head of $d_i$,
   respectively \COMMENT{ \vspace{5pt} \\ \small \% reduce
    the capacities of $d$ and $\rev(d)$ by $M$ and adjust the flow
    appropriately}
   \FOR{$d \in \set{ d_i, \rev(d_i)}$} \label{0-loop-drevd}
      \STATE $\cvec[d] := \cvec[d] - M$ \label{0-cap}
      \STATE $\fvec'[d] := \min \set{\fvec'[d], \cvec[d]}$
      \STATE $\fvec'[\rev(d)] := - \fvec'[d]$\label{0-reduce-flow}
\COMMENT{\vspace{5pt}}
   \ENDFOR
   \STATE {\it excess} $:=$ inflow at $p_i$

   \STATE \algorithmicif { $ {\it excess} > 0$} \algorithmicthen \ $d
   := d_i$
   \algorithmicelse \ $d := \rev(d_i)$ \label{0-test-inflow}
   \COMMENT{find in which direction flow should be pushed}
   \STATE add to $\fvec'$ a maximum feasible flow from $\tail(d)$ to $\head(d)$ with limit {\it excess}  \label{0:flow}
\ENDFOR
\RETURN $\fvec'$
\end{algorithmic}
\end{algorithm}

\subsection{An Inefficient Implementation}

In this section, we give 
an {\em inefficient} implementation of line~\ref{0:flow} of the abstract algorithm. This will facilitate the explanation of the efficient procedure in the next section.
 We first review the necessary ideas and tools.

\subsubsection{Hassin's algorithm for maximum $st$-planar flow} \label{sec:st-planar}
An $st$-planar graph is a planar graph in which nodes $s$ and $t$ are incident to the same face.
Hassin~\cite{Hassin81} gave an algorithm for computing a maximum flow
from $s$ to $t$ in an $st$-planar graph. We briefly describe this
algorithm here since we use it in
implementing line~\ref{0:flow} of Algorithm~\ref{alg:fix}.

Hassin's algorithm starts by adding to $G$ an artificial infinite
capacity arc $a$ from $t$ to $s$. 
Let $d$ be the dart that corresponds to $a$ and whose head is $t$.
Let $t^*$ be the head in $G^*$ of the dual of $d$. 
Compute in the dual $G^*$ a shortest path tree rooted at $t^*$, where
the length of a dual dart is defined as the capacity of
the primal dart.
Let $\phivec[\cdot]$ denote the shortest path distances from $t^*$ in
$G^*$. 
Consider the flow 
\begin{equation}
\rhovec[d'] = \phivec[\head_{G^*}(d')] -
\phivec[\tail_{G^*}(d')] \mbox{for all darts $d'$}\label{eq:circ-pot}
\end{equation}
After removing the artificial arc $a$ from $G$, $\rhovec$
is a maximum feasible flow from $s$ to $t$ in $G$.
We say that $\phivec$ is a \emph{face potential}
vector that induces $\rhovec$.

In our algorithm we are interested in a {\em max flow with limit} $x$ from $s$ to
$t$ rather than a maximum flow, i.e., a flow
whose value is at most a given number $x$ but is otherwise maximal. 
It is not difficult to see that setting the capacity of the artificial arc $a$
to $x$ instead of infinity results in the desired limited max flow~\cite{KNK93}.

In our implementation, instead of using an artificial arc from $t$ to $s$,
we use an existing arc whose endpoints are $s$ and $t$ as the arc $a$ above.
In order for this to work, the capacity of the dart $d$ that
corresponds to $a$ and whose head is $t$ must be zero (as is indeed the case
if $a$ is an artificial arc from $t$ to $s$). This can always be
achieved by first pushing flow on $d$ to saturate it.
Also note that in this case, we do not remove $a$ from $G$. 
Hence, $\rhovec$ is a feasible circulation, rather than a maximum
flow, since flow is being pushed back
from $t$ to $s$ along $a$. To convert it into a maximum flow one just
has to undo the flow on $a$. If we define $\fvec$ by
\begin{equation}
\fvec[d'] = \left\{ \begin{array}{ll}
-\rhovec[d'] & \mbox{if $ d' $ corresponds to $a$} \\
0  & \mbox{otherwise}
\end{array} \right. , 
\end{equation}
then $\rhovec + \fvec$ is a maximum feasible $st$-flow. We will use
the fact that this flow can be represented implicitly by the face potential
vector $\phivec$, and the flow values $\fvec[d']$ for the two darts
corresponding to $a$.

\subsubsection{The Inefficient Implementation}
Recall that $\fvec_0$ is the flow at the beginning of the procedure. 
Observe that the change to the flow in iteration~$i$ of the abstract algorithm
(line~\ref{0:flow}) is a flow between the endpoints of the dart $d_i$.
As discussed in Section~\ref{sec:st-planar}, this 
 flow can be represented as the sum of ($i$) a circulation
$\rhovec_i$ and ($ii$) a flow on $d_i$ and $\rev(d_i)$.
Summing over the first $i$ iterations,
the flow $\fvec'$ at that time can be represented as the sum 
\begin{equation} \label{eq:flow-form} 
\fvec' = \rhovec + \fvec
\end{equation}
where
$\rhovec = \sum_{j=1}^i\rhovec_j$
is a circulation and $\fvec$ is a flow
assignment that differs from $\fvec_0$ only on the darts of
$\set{d_j}_{j=1}^i$ and their reverses. 

The inefficient implementation of line~\ref{0:flow} of the abstract algorithm appears as
Algorithm~\ref{alg:inefficient-fix}.  We now describe it.
The total flow $\fvec'$ is maintained by representing $\fvec$ and the circulation $\rhovec$ as in Eq.~\eqref{eq:flow-form}. 
$\fvec$ is represented explicitly, but, in preparation for the efficient
implementation, the circulations
$\rhovec_j$  are represented implicitly by the face-potentials $\phivec_j$.
By linearity of Eq.~\eqref{eq:circ-pot},  the sum 
$\phivec = \sum_{j=1}^i\phivec_j$ is the face potential vector that induces the circulation $\rhovec$.

Recall that $d$ is the dart of $C$ such that flow
needs to be sent from $\tail(d)$ to $\head(d)$ (line~\ref{0-test-inflow} of Algorithm~\ref{alg:fix}).
In lines~\ref{1-residual-capacity} -- \ref{1-push-on-revd}, the procedure pushes as much as possible on $d$
itself. Consequently, either
$d$ is saturated or conservation at $p_i$ is achieved.

Next, an implementation of Hassin's algorithm pushes a maximum flow with limit
$|\inflow(p_i)|$ from $\tail(d)$ to
$\head(d)$.
The procedure first sets the length of darts to their residual capacities
(line~\ref{1-residual-capacities}) and sets the 
length of $\rev(d)$ to be the flow limit $|\inflow(p_i)|$ (line~\ref{1-bound}). 
Since the flow maintained is feasible, all residual capacities are non-negative.
The procedure
then computes all the $\head_{G^*}(d)$-to-$f$ distances $\phivec_i[f]$ in $G^*$ using Dijkstra's algorithm
(line~\ref{1-dijkstra}).  
Let $\rhovec_i$ be the circulation
corresponding to the face-potential vector $\phivec_i$.
The procedure sets {\it val} equal
to $\rhovec[d]$ in line~\ref{1-flow-value}, then
subtracts {\it
  val} from $\fvec[d]$ and adds it to $\fvec[\rev(d)]$.  Finally, in the last
line, the current
circulation is added to the accumulated circulation by adding the potential
$\phivec_i$ to
$\phivec$.

\begin{algorithm}[h!]\caption{Inefficient Implementation of
    line~\ref{0:flow} of \proc{AbstractFixConservationOnPath} (Algorithm~\ref{alg:fix})}
\label{alg:inefficient-fix}
\begin{algorithmic}[1]
\vspace{15pt}
   \STATE {\it residual capacity} $:= \cvec[d] - \fvec[d] -\phivec[\head_{G^*}(d)] + \phivec[\tail_{G^*}(d)]$\label{1-residual-capacity}\COMMENT{\vspace{-23pt}\\\hspace{-2pt} \small \% push flow on $d$ to make
    its residual capacity zero as required for Hassin's algorithm \vspace{11pt}}
   \STATE {\it val} $:= \min \set{\text{\it residual capacity},
     |\inflow(p_i)|}$ \COMMENT{ amount of flow to push on $d$} \label{1-amount-to-push-on-d}
   \STATE $\fvec[d] := \fvec[d] +  \text{\it val}$ ; $\fvec[\rev(d)]
   := -\fvec[d]$ 
\label{1-push-on-revd}
   \COMMENT{\vspace{5pt}\\ \small \% push excess inflow from $\tail(d)$ to $\head(d)$
     using Hassin's algorithm}
   \STATE let $\ell[d'] := \cvec[d'] - \fvec[d']
   -\phivec[\head_{G^*}(d')] + \phivec[\tail_{G^*}(d')]$ for all darts $d'
   \in G$ \label{1-residual-capacities}    \COMMENT{lengths are residual capacities}
   \STATE $\ell[\rev(d)] := |\inflow(p_i)|$ \label{1-bound} \COMMENT{
     set the limit on the residual capacity of $\rev(d)$}
   \STATE $\phivec_i(\cdot) := \proc{Dijkstra}(G^*, \ell,
   \head_{G^*}(d))$ \label{1-dijkstra} \COMMENT{face
     potential are  distances in $G^*$ from $\head_{G^*}(d)$ w.r.t.\ residual capacities}
   \STATE {\it val} $:= \phivec_i[\head_{G^*}(d)] -
   \phivec_i[\tail_{G^*}(d)]$ \COMMENT{ the amount of flow
     assigned to $d$ by the circulation
     corresponding to $\phivec_i$} \label{1-flow-value}
   \STATE $\fvec[d] := \fvec[d] - \text{\it val}$ ; $\fvec[\rev(d)] :=
   -\fvec[(d)]$\label{1-not-on-revd2} \COMMENT{ do not push the
     circulation on $d$ and $\rev(d)$}
   \STATE $\phivec  = \phivec +
   \phivec_i$ \label{1-accumulate}\COMMENT{ accumulate the current
     circulation}
\end{algorithmic}
\end{algorithm}
\subsection{An Efficient Implementation}\label{sec:eff}
In this section we give 
an {\em efficient} implementation of Algorithm~\ref{alg:fix}. 
We first review the necessary tools.
\subsubsection{Fakcharoenphol and Rao's Efficient Dijkstra Implementation} \label{sec:FR-Dijkstra}
Let $X$ be a set of nodes. Let $H$ be a planar graph in which the nodes of $X$ lie a single face.
Let $x_1, x_2, \cdots$ be the clockwise order of the nodes of $X$ on that
face.
Let $P$ be a set of darts not necessarily in the graph $H$ whose endpoints are
nodes in $X$.
Fakcharoenphol and Rao~\cite{FR06} described a data structure that can be
used in a procedure that efficiently implements Dijkstra's algorithm in
$H \cup P$. The
 procedure takes as input a
table $\DIST$ such that $\DIST[i,j]$ stores the distance between $x_i$
and $x_j$ in $H$, an array $\ell$ that stores the lengths of the darts
in $P$,
 and a node $v \in X$. It is assumed that the lengths in $\DIST$ and
 in $\ell$ are non-negative.
The procedure outputs the distances of the nodes of $X$ from $v$ in $H \cup
P$ in $O(|X| \log^2 |X| + |P| \log|X|)$-time.

We mention a technical issue whose importance will become apparent in
the sequel. The procedure partitions the table $\DIST$ into several subtables
$\{\DIST_\alpha\}_\alpha$ that correspond to distances between pairs of disjoint
sets of nodes of $X$, where each set consists of nodes that are consecutive
in $X$. It is assumed in~\cite[footnote on p. 884]{FR06} that for each such subtable $\DIST_\alpha$, a data structure 
that supports range minimum queries of the form $\min_{j_1 \leq j \leq j_2} 
\set{\DIST_\alpha[i,j]}$ for every $i,j_1,j_2$ is given. Fakcharoenphol and Rao 
note that such a data structure can be easily implemented by
using a range-search tree~\cite{DVOS00} for every row $i$ of $\DIST_\alpha$. The
time required to construct  all of the range-search trees for
$\DIST_\alpha$ is proportional to the size of $\DIST_\alpha$.

\subsubsection{Price Functions, Reduced Lengths and FR-Dijkstra} \label{sec:price}

For a directed graph $G$ with dart-lengths $\ell(\cdot)$,
a {\em price function} is a function $\phi$ from the nodes of $G$ to
the reals.  For a dart $d$, the {\em reduced length with respect to
  $\phi$} is $\ell_\phi(d) = \ell(d) +\phi(\tail(d))-\phi(\head(d))$.
A {\em feasible} price function is a price function that induces
nonnegative reduced lengths on {\em all} darts of $G$ (see ~\cite{J1977}).


Single-source distances form a feasible price function.
Suppose that, for some node $r$ of $G$, for
every node $v$ of $G$, $\phi(v)$ is the $r$-to-$v$ distance in $G$
with respect to $\ell(\cdot)$.  Then for every arc $uv$, $\phi(v)
\leq \phi(u) + \ell(uv)$, so $\ell_\phi(uv) \geq 0$. Here we assume,
without loss of generality, that all distances are finite (i.e., that
all nodes are reachable from $r$) since we can always add arcs with
sufficiently large lengths to make all nodes reachable without
affecting the shortest paths in the graph.

We will use the following variant of Fakcharoenphol and Rao's efficient Dijkstra
implementation. 
The procedure \proc{FR}$(\DIST , \ell, \phivec^X, v)$ takes as input the 
table $\DIST$ and the array $\ell$ as described above, as well as 
a feasible price function $\phivec^X$ on the nodes of $X$ and a node $v \in
X$. It outputs the distances of the nodes of $X$ from $v$ in $H \cup
P$ w.r.t.\ the reduced lengths w.r.t.\ $\phivec^X$. We stress that lengths in
$\DIST$ and in $\ell$ may be negative, but the reduced lengths are all
non-negative. The computation
takes $O(|X| \log^2 |X| + |P| \log|X|)$ time.

This differs from the procedure described in Section~\ref{sec:FR-Dijkstra} 
only in the existence of the price function $\phivec^X$.
We cannot afford to compute the entire table of reduced
lengths since that would dominate the running time of the algorithm in~\cite{FR06}.
Instead,  whenever their algorithm requires some specific reduced
length, we can compute it in constant time from $\DIST$.
This, however, does not suffice. 
Recall that the algorithm in~\cite{FR06} requires that, for each
of the subtables $\DIST_\alpha$,
 range-search trees that support range minimum queries of the form $\min_{j_1 \leq j \leq j_2}
\set{\DIST_\alpha[i,j] + \phivec^X[x_i] - \phivec^X[x_j]}$ for every $i,j_1,j_2$ are given.
Note that the results of such queries may be different
for different price functions. Computing the range-search trees would
take $O(|X|^2)$ which will dominate the running time of the entire
procedure. To overcome this difficulty we use Monge
range-query data structures, due to Kaplan and
Sharir~\cite{KS11}, which can be constructed from
the table $\DIST$ in $O(|X| \log |X|)$ time, and answer queries of the
desired form in $O(\log |X|)$ time. 

\subsubsection{The Procedure}
Finally, we describe the efficient implementation.  
The procedure
keeps track of the inflow at each node of $P$ in an array
$\vvec[\cdot]$.
As in the inefficient implementation, the procedure will
maintain the total flow as the sum of a circulation $\rhovec$ and a
flow assignment $\fvec$ that differs from $\fvec_0$ only on the darts
of $P\cup \rev(P)$. Initially $\fvec$ is set equal to $\fvec_0$. The circulation $\rhovec$ will be represented by a
face-potential vector $\phivec$.  However, we will show that it suffices to
maintain just those entries of $\phivec$ that correspond to faces incident
to $P$.

\begin{figure}
  \begin{center}
    \begin{minipage}[c]{0.40\linewidth}
      \includegraphics[scale=0.45]{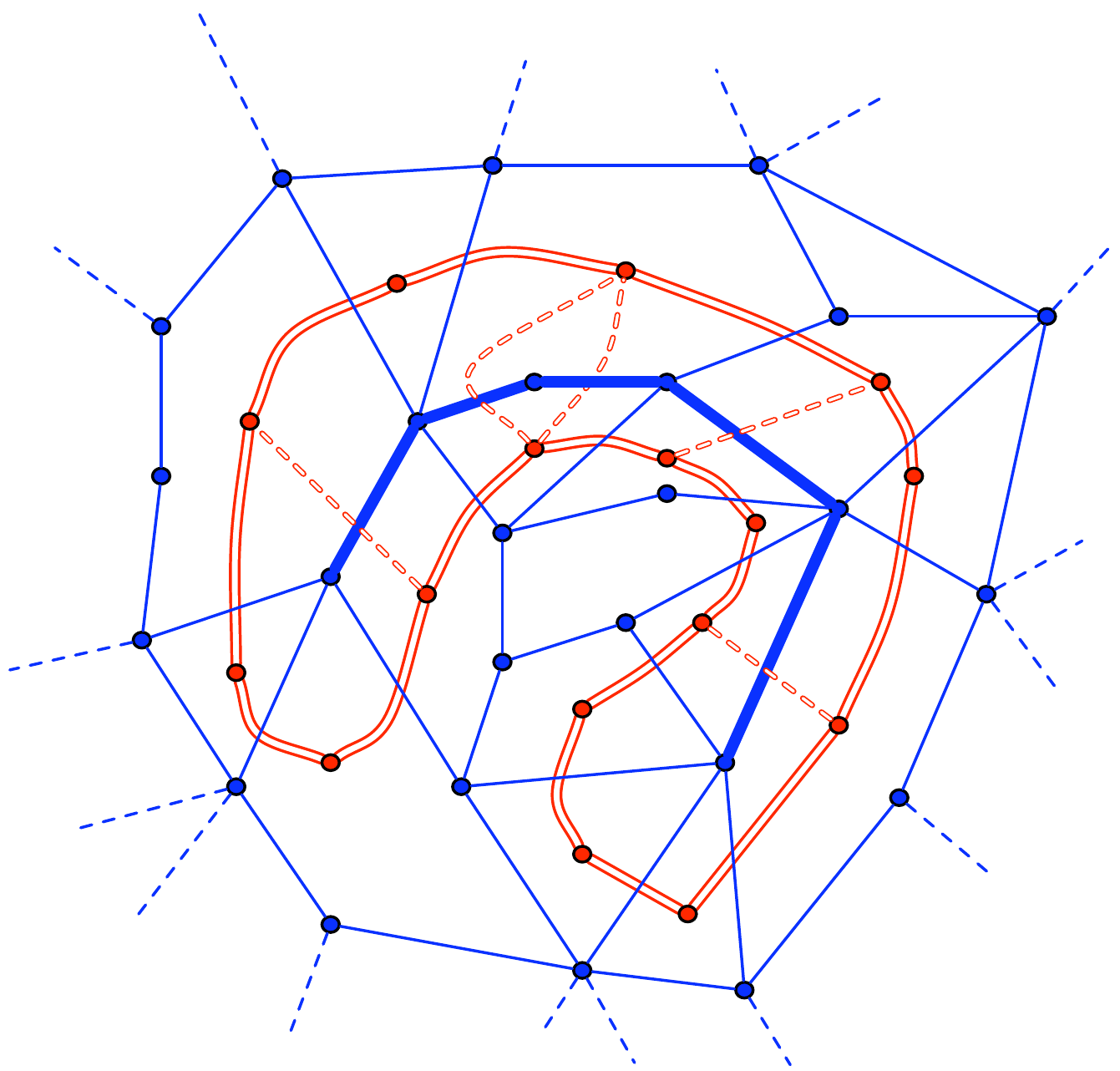}
    \end{minipage}\hfill
    \begin{minipage}[c]{0.56\linewidth}
\caption{An example illustrating that the nodes of $X^*$ are on the boundary of a single face of $H^*$. The diagram shows part of the graph $G$ and some edges of its dual $G^*$. Edges of $G$ are solid blue. Edges of $P$ are bold. Dual edges are double lined red. The dual edges of $P$ are in double lined dashed red.}
\label{fig:singleface}
    \end{minipage}
  \end{center}
\end{figure}

Define each dart $d$'s length by $\ell(d)=\cvec[d] - \fvec[d]$.
Let $X^*$ be the set of endpoints in the planar dual $G^*$ of the
darts of $P$ (i.e. the primal faces incident to $P$).
Let $H^*$ be the graph obtained from $G^*$ by removing the darts of
$P$. Note that in $H^*$, all the nodes of $X^*$ that did not disappear
(i.e., that have degree greater than zero) are on the boundary of a
single face, see Figure~\ref{fig:singleface}.

The procedure uses a multiple-source shortest paths (MSSP) algorithm~\cite{Klein05,CabelloC07} to compute a table $\DIST^*[\cdot,\cdot]$
of $X^*$-to-$X^*$ distances in $H^*$ with respect to the lengths $\ell(\cdot)$. 
 
The implementation of lines~\ref{0-1}--~\ref{0-test-inflow} of the abstract algorithm using the chosen representation of $\fvec'$ is straightforward.
We therefore focus on the implementation of line~\ref{0:flow} of
Algorithm~\ref{alg:fix}, given as
Algorithm~\ref{alg:efficient-fix}.
 
Consider iteration $i$ of the algorithm. 
The main difference between the inefficient implementation and the efficient one is in implementing the
Dijkstra step for computing shortest paths in the dual.
Instead of computing the entire shortest-path tree, the procedure computes just the distance
labels of nodes in $X^*$. This is done using the FR data structure,
whose initialization requires the $X^*$-to-$X^*$ distances in $H^*$  with respect to the 
residual capacities. We now explain how these distances can be
obtained.

The flow on a dart $d$ in the primal is 
\begin{equation}
\fvec[d] + \phivec[\head_{G^*}(d)] -\phivec[\tail_{G^*}(d)].
\end{equation}
Therefore the residual capacity of $d$ is 
\begin{equation} \label{eq:residual-capacity}
(\cvec[d] - \fvec[d]) - \phivec[\head_{G^*}(d)] + \phivec[\tail_{G^*}(d)]
\end{equation}
which is its {\em reduced} length $\ell_{\phivec}[d]$ with respect to
the length $\ell(\cdot)$ and price function $\phivec$.

Suppose that $d$ belongs to $H^*$, i.e. $d$ is not one of the darts of
$P\cup \rev(P)$.  The procedure never changes $\fvec[d]$, so
$\fvec[d]=\fvec_0[d]$.  Therefore $\ell(d) = \cvec[d]-\fvec_0[d]$.
These lengths are known at the beginning of the procedure's execution.
The reduced length of an $X^*$-to-$X^*$ path $Q=d'_1,d'_2,
\cdots, d'_{j}$ in $H^*$ is
\begin{equation}
\sum_{i=1}^j \left( \ell(d'_i) -\phivec[\head_{G^*}(d'_i)] +
\phivec[\tail_{G^*}(d'_i)] \right) = 
\left( \sum_{i=1}^j \ell(d'_i) \right) -\phivec[\pend(Q)] +
\phivec[\pstart(Q)]. \label{eq:Q-red-len}
\end{equation}
Therefore, for any nodes $x,y\in X^*$, the $x$-to-$y$ distance in
$H^*$ w.r.t.\ the residual capacity is given by
$\DIST^*[x,y]-\phivec[y]+\phivec[x]$.  Since the procedure maintains
the restriction of $\phivec$ to faces of $X^*$, this distance can be
obtained in constant time.
Adapting FR's data structure to handle reduced distances w.r.t.\ a
price function $\phivec$ was discussed in Section~\ref{sec:FR-Dijkstra}.

\begin{algorithm}[h!]\caption{Efficient Implementation of line~\ref{0:flow} of \proc{AbstractFixConservationOnPath} (Algorithm~\ref{alg:fix})}
\label{alg:efficient-fix}
\begin{algorithmic}[1]
\vspace{15pt}
   \STATE {\it residual capacity} $:= \cvec[d] - \fvec[d] -\phivec^X[\head_{G^*}(d)] + \phivec^X[\tail_{G^*}(d)]$\label{2-residual-capacity}\COMMENT{\vspace{-23pt}\\\hspace{-2pt} \small \% push flow on $d$ to make
    its residual capacity zero as required for Hassin's algorithm \vspace{11pt}}

   \STATE {\it val} $:= \min \set{\text{\it residual capacity}, |\vvec[p_i]|}$ \COMMENT{ amount of flow pushed on $d$}
   \STATE $\fvec[d] = \fvec[d] +  \text{\it val}$ ; $\fvec[\rev(d)]
   := -\fvec[d]$ 
\label{2-push-on-revd}
   \STATE $\vvec[\tail(d)] = \vvec[\tail(d)] - \text{\it val}$ ; $\vvec[\head(d)]  = \vvec[\head(d)] +   \text{\it val}$ \COMMENT{ update the inflow at $p_i$ and $p_{i+1}$
\vspace{6pt}\\ \small \small \% push excess inflow from $\tail(d)$ to $\head(d)$
     using Hassin's algorithm}
   \STATE let $\ell[d'] := \cvec[d'] - \fvec[d']$ for all darts $d'
   \in P\cup \rev(P)$ \label{2-residual-capacities} \COMMENT{\parbox{2.4in}{lengths
     of explicit darts are residual
     capacities (not including circulation component) }} \vspace{2pt}
   \STATE $\ell[\rev(d)] := |\vvec[p_i]|
   -\phivec^X[\tail_{G^*}(\rev(d))] + \phivec^X[\head_{G^*}(\rev(d))]$
   \COMMENT{ \parbox{2.2in}{set the limit on the residual capacity of $\rev(d)$
     (adjusted by circulation component)}} \vspace{2pt}
   \STATE $\phivec_i^X(\cdot) := FR(\DIST^* , \ell, \phivec^X,
   \head_{G^*}(d))$ \label{2-FR}
\COMMENT{\parbox{3in}{face potential are distances in $G^*$ from $\head_{G^*}(d)$
  w.r.t.\  the reduced lengths induced by $\phivec^X$} }
   \STATE {\it val} $:= \phivec_i^X[\head(d)] - \phivec_i^X[\tail(d)]$\label{2-val} \COMMENT{the amount of flow
     assigned to $d$ by the circulation  corresponding to $\phivec_i^X$}
   \STATE $\fvec[d] := \fvec[d] - \text{\it val}$ ; $\fvec[\rev(d)] :=
   -\fvec[(d)]$\label{2-not-on-revd2} \COMMENT{ do not push the
     circulation on $d$ and $\rev(d)$}
   \STATE $\phivec^X  = \phivec^X + \phivec_i^X$ \COMMENT{ accumulate the current circulation}
   \STATE $\vvec[\tail(d)] = \vvec[\tail(d)] - \text{\it val}$ ;
   $\vvec[\head(d)]  = \vvec[\head(d)] +   \text{\it val}$ \COMMENT{
     update the inflow at $p_i$ and $p_{i+1}$}
\end{algorithmic}
\end{algorithm}

It follows from the above discussion that, after executing 
line~\ref{0:flow} of Algorithm~\ref{alg:fix} using the efficient
implementation (Algorithm~\ref{alg:efficient-fix}) for the last time, 
the flow assignment $\fvec$ maintained by the efficient implementation is the same as the
one that would have been computed if the inefficient implementation
(Algorithm~\ref{alg:inefficient-fix}) were used. Moreover, the potential
function $\phivec^X$ computed by the efficient implementation is the restriction to
$X^*$ of the potential function $\phivec$ that would have been computed by the inefficient implementation.

In order to output the flow $\fvec'$ in the entire graph $G$, we need to know
the potential function $\phivec$ rather than just its restriction
$\phivec^X$. Observe, however, that any pseudoflow that differs from
$\fvec$ by a circulation 
is a valid output of \proc{FixConservationOnPath} since a circulation does not change the inflow at
any node, nor does it introduce residual paths between nodes that are
not connected by a residual path in $G_{\fvec}$. 
It therefore suffices to find any feasible circulation $\chivec$ in $G_{\fvec}$. 
This can be done by computing shortest paths in $G^*$ w.r.t.\ the
lengths $\ell := \cvec - \fvec$ from an arbitrary node $x \in G^*$. 
Note, however, that for darts of $P \cup \rev(P)$ these lengths might be 
negative. We therefore use the $O(n \log^2n / \log\log n)$-time
algorithm for shortest paths with negative lengths in planar
graphs~\cite{MWN10} to compute a feasible circulation $\chivec$. 
~\footnote{In appendix~\ref{sec:alternative}, we show how to use one more call to \proc{FR} followed
by a call to Dijkstra's algorithm to compute these distances. While we
think doing so is more elegant and simpler to implement, it does not
change the asymptotic running time of the algorithm.}

The pseudocode for the efficient implementation of
\proc{FixConservationOnPath} is given in appendix~\ref{sec:fix-code}.

\paragraph{Running Time Analysis}
Let $n$ and $k$ be the number of nodes in $G$ and in $P$, respectively.
The initialization time is dominated by the $O(n\log n + k^2\log n)$ time for MSSP.
The execution of each iteration of the main loop is dominated by the
call to \proc{FR}, which takes $O(k \log^2 k)$ time.
The number of iterations is $k-1$, leading to a total of $O(k^2
\log^2 k)$ time for execution of the main loop.
Computing the circulation $\chivec$ requires one 
shortest path computation,
which takes $O(n \log^2n / \log\log n)$ time.
Thus total running time of the efficient implementation of
\proc{FixConservationOnPath} is $O(n \log^2 n / \log\log n + k^2 \log^2 k)$.

\section{Pushing Excess Inflow from a Cycle 
}\label{sec:C2A}
In this section we describe the procedure
\proc{CycleToSingleSinkLimitedMaxFlow} that pushes excess inflow from a
cycle to a node not on the cycle. The procedure is given as Algorithm~\ref{alg:lim_cyc}.
 The procedure \proc{SingleSourceToCycleLimitedMaxFlow} is very similar. We omit its description.

\begin{algorithm}\caption{ \proc{CycleToSingleSinkLimitedMaxFlow}($G,\cvec, \fvec_0, C, t$) \label{alg:lim_cyc}}
{\bf Input:} a directed planar graph $G$ with capacities $\cvec$, a pseudoflow $\fvec_0$, a simple cycle $C$, a sink $a_i$. \\
{\bf Assumes:}  $\forall v \in C^+,\ \inflow_{\fvec_0}(v) \geq 0$, 
 where 
$C^+=\{ v \in C : \{x\in C : \inflow_{\fvec_0} (x)>0\} \stackrel{G_{\fvec_0}}{\rightarrow} v\}$.\\
{\bf Output:} a pseudoflow $\fvec$ s.t.
{\it(i)} $\fvec - \fvec_0$ obeys conservation everywhere but
$C^+ \cup \{t\}$, \\
{\it(ii)} $\forall v \in C^+,\ \inflow_{\fvec}(v) \geq 0$, 
{\it (iii)} $\{ v \in C : \inflow_{\fvec}(v) > 0\}
\stackrel{G_{\fvec}}{\nrightarrow} a_i$.\\
\begin{algorithmic}[1]
\STATE \vspace{-10pt}let $C^+ := \{ v \in C : \textrm{there exists a residual path to $v$ from a node } x\in C \textrm{ with } \inflow_{\fvec_0} (x)>0 \}$ \label{CTS:C+}
\STATE delete the nodes of $C \setminus C^+$\label{CTS:del}
\STATE let $v_1,v_2, \dots, v_{\ell}$ be the nodes of $C^+$, labeled
according to their cyclic order on $C$
\STATE add artificial arcs $v_iv_{i+1}$ for $1 \leq i < \ell$\label{CTS:add}
\STATE let $P$ be the $v_1$-to-$v_\ell$ path of artificial arcs
\STATE contract all the edges of $P$, collapsing $C^+$ into the single node $v_1$\label{CTS:contract}
\STATE $\fvec :=$
\proc{SingleSourceSingleSinkMaxFlow}($G,\cvec-\fvec_0,c_1,t$) \label{CTS:stmaxflow}
\STATE undo the contraction of  the edges of $P$\label{CTS:uncontract}
\STATE $\fvec := $
\proc{FixConservationOnPath}($G,P,\cvec,\fvec_0+\fvec$) $- \fvec_0$\label{CTS:fix}
\STATE modify $\fvec$ to push back flow to nodes of $C^+$ whose inflow
w.r.t $\fvec_0+\fvec$ is negative\label{CTS:pushback}
\STATE $\fvec := \fvec_0 + \fvec$
\RETURN $\fvec$
\end{algorithmic}
\end{algorithm}

To compute $C^+$ in Line~\ref{CTS:C+}, consider the residual graph of
$G$ w.r.t.\ $\fvec_0$. Add a node $v$ and non-zero capacity arcs $vw$
for every node $w$ whose inflow w.r.t.\ $\fvec_0$ is positive (these
arcs may not preserve planarity). In $O(|G|)$ time, find the set $X$ of nodes that are reachable from $v$ via darts with non-zero capacity. Then $C^+ = C\cap X$. 

Since $C^+$ consists of all nodes of $C$ reachable via
residual paths from the
nodes of $C$ with positive inflow, the flow computed by the procedure
involves no darts incident to nodes in $C \setminus
C^+$. Thus, restricting the computation to the graph obtained
from $G$ by deleting the nodes in $C \setminus C^+$ (line~\ref{CTS:del}) does not restrict
the computed flow. After deletion, adding artificial arcs between
consecutive nodes of $C^+$ (line~\ref{CTS:add}) will not violate planarity. 
Contracting the artificial arcs effectively turns $C^+$ into a single node $v_1$.
Next,  the procedure computes a maximum flow $\fvec$ from $C^+$ to $t$
w.r.t.\ residual capacities $\cvec - \fvec_0$. This is done by invoking a single-source single-sink maximum flow
algorithm~\cite{BorradaileK09} with source $v_1$ and sink
$t$ (line~\ref{CTS:stmaxflow}).
Uncontracting the artificial arcs turns $\fvec$
into a
maximum $C^+$-to-$t$ flow in $G$ w.r.t.\ the residual capacities $\cvec
- \fvec_0$. However, some of the nodes of $C^+$ may have negative
inflow w.r.t.\ $\fvec_0 + \fvec$. In line~\ref{CTS:fix}, the procedure
calls \proc{FixConservationOnPath} to reroute the flow $\fvec$ among the
nodes of $C^+$ so that, w.r.t.\ $\fvec_0 + \fvec$, there are no
residual paths from nodes of $C^+$ with positive inflow to nodes of
$C^+$ with negative inflow. This implies that any node of $C^+$ that
still has negative inflow has pushed too much
flow. Line~\ref{CTS:pushback} modifies $\fvec$ to push back such
excess flow so that
no node of $C^+$ has negative inflow w.r.t.\ $\fvec_0 + \fvec$. This is
done using the procedure of Section~\ref{sec:prel}
described in the appendix.
Finally, the procedure returns $\fvec_0+\fvec$. See Appendix~\ref{apn:c2a-cor} for a formal proof of correctness.

We next analyze the running time of this procedure on an $n$-node
graph $G$ and a $k$-node
cycle $C$. The $st$-maximum flow computation in
line~\ref{CTS:stmaxflow} takes $O(n\log n)$ time using the
algorithm of Borradaile and Klein~\cite{BorradaileK09}.
The running time of the procedure is therefore dominated by the call to  \proc{FixConservationOnPath}
in line~\ref{CTS:fix} which takes $O(n \log^2 n / \log\log n + k^2 \log^2 k)$ time.

\section*{Acknowledgements}
We thank Haim Kaplan and Micha Sharir for 
discussions of their unpublished data structure~\cite{KS11}. 
PNK and SM thank Robert Tarjan for encouraging us to work on this problem.

\bibliographystyle{plain}
\bibliography{msms2}
\newpage
\begin{appendix}

\section{Correctness of Algorithm~\ref{alg:msms}}\label{sec:correct}

\begin{lemma}\label{lem:circ-no-res}
Let $\rhovec$ be a circulation. Let $u$ and $v$ be nodes in a graph
$G$. Then
$$u \stackrel{G} \nrightarrow v \Rightarrow u \stackrel{G_{\rhovec}} \nrightarrow v.$$
\end{lemma}
\begin{proof}
Pushing a circulation does not change the amount of flow crossing any
cut. This implies that if there was no
$u$-to-$v$ residual path before $\rhovec$ was pushed, then there is none
after $\rhovec$ is pushed. 
\end{proof}

We will use the following two lemmas in the proof of correctness:
\begin{lemma}(sources lemma)\label{lem:suf}
Let $f$ be a flow with source set $X$. 
Let $A,B$ be two disjoint sets of nodes. Then
$$A \cup X \stackrel{G} \nrightarrow B \Rightarrow A \stackrel{G_f} \nrightarrow B.$$
\end{lemma}
\begin{proof}
The flow $f$ may be decomposed into a cyclic component (a circulation) and an
acyclic component.
By Lemma~\ref{lem:circ-no-res}, it suffices to show the lemma for an acyclic flow $f$.

Suppose for the sake
of contradiction that there exists
a residual $a$-to-$b$ simple path $P$ after $f$ is pushed for some $a\in A$
and $b \in B$.
Let $P'$ be the maximal suffix of $P$ that was residual before the
push.  Maximality implies that the dart $d$ of $P$ whose head is $\pstart(P')$ was
non-residual before the push, and $f(\rev(d)) >
0$. The fact that $f(\rev(d)) > 0$ implies that before $f$ was pushed
there was a residual path $Q$ from some node $x \in X$ to
$\head(d)$. Therefore, the concatenation of $Q$ and
$P'$ was a residual
$x$-to-$b$ residual path before the push, a contradiction.
\end{proof}

The proof of the following lemma is symmetric.
\begin{lemma}(sinks lemma)\label{lem:pref}
Let $f$ be a flow with sink set $X$. 
Let $A,B$ be two disjoint sets of nodes. Then
$$A \stackrel{G} \nrightarrow B \cup X \Rightarrow A \stackrel{G_f} \nrightarrow B.$$
\end{lemma}
For node sets
$W,Y,Z$, we will use the notation 
\sources$(W,Y,Z)$ to indicate the use of the sources lemma to
establish that there are no $W$-to-$Y$ residual paths after a flow
with source set $Z$ is pushed.
We will use \sinks$(W,Y,Z)$ in a similar manner.


\begin{lemma}\label{lem:between-recurse-and-pseudo}
At any time in the running of the algorithm after the last execution of
line~\ref{line:recurse} and before the execution of line~\ref{line:pseudo},
$S\nrightarrow T $, $S \nrightarrow A$, $S \nrightarrow C$, $A \nrightarrow
T$, $C \nrightarrow T$.
\end{lemma}
\begin{proof}
The fact that the properties hold just after line~\ref{line:recurse}
follows from the properties of the recursive calls and from the fact
that any residual path from a node in one piece to a node in the other
consists of a residual path to $C$ and a residual path from $C$. 
Note that, because of the contractions in line~\ref{line:contract}, at
this time the cycle $C$ consists of just the node $v$. The the nonexistence of residual paths
with respect to $v$ in the recursive calls implies the nonexistence of residual paths w.r.t.\ any
node of $C$ after the contractions are undone in line~\ref{line:uncontract}. 

Now, let us prove that the properties are maintained until just before
the execution of line~\ref{line:pseudo}.
By the above argument, there is a cut separating $S$ from $T\cup A\cup C$ 
which is saturated just after line~\ref{line:uncontract}.
The procedure in line~\ref{line:fix} only pushes flow between vertices
of $C$, 
and in lines~\ref{line:i-loop}--~\ref{line:from-ai}, flow is only pushed
between the nodes of $A$ and $C$. These sets are all on the same side of the cut
which therefore stays saturated. It follows that $S\nrightarrow T $, $S \nrightarrow A$,
and $S \nrightarrow C$ at any point after line~\ref{line:recurse} and before
line~\ref{line:pseudo}. A similar argument applied to a saturated cut
separating 
$A\cup C$ from $T$ shows $A \nrightarrow T$ and $C \nrightarrow T$.
\end{proof}

Recall that $C_i^+$ is the set of nodes of $C$ that are
reachable via residual paths from some node of $C$ with positive excess at the
beginning of iteration $i$, and that $C_i^-$ is the set of nodes of $C$ that have residual
paths to some node of $C$ with negative excess at the beginning of iteration $i$.
\begin{lemma}
Just after line~\ref{line:fix} of Algorithm~\ref{alg:msms} is executed, $C_1^+ \nrightarrow C_1^-$.
\end{lemma}
\begin{proof}
Follows from the definition of \proc{FixConservationOnPath}.
\end{proof}

\begin{lemma}
For all $i<j$, $C^+_j \subseteq C_i^+$ and $C^-_j \subseteq C_i^-$
\end{lemma}
\begin{proof}
It suffices to show that, for all $i$, $C^+_{i+1} \subseteq C^+_i$ and $C^-_{i+1}
\subseteq C^-_i$.

The flow pushed in iteration $i$ of line~\ref{line:to-ai} can be decomposed into a flow whose sources and
sinks are all in $C^+_i$ and a flow whose sources are in $C_i^+$ and
whose sink is $a_i$. Therefore, the set $X$ of nodes of $C$ with positive
inflow immediately after iteration $i$ of line~\ref{line:to-ai} is a
subset of $C^+_i$. 
By definition of \proc{SingleSourceToCycleLimitedMaxFlow}, the set of
nodes of $C$ with positive inflow does not change after
line~\ref{line:from-ai} is executed.
Therefore, $C^+_{i+1}$ is the set of nodes reachable from $X$ by a residual
path after iteration $i$ of line~\ref{line:from-ai}.

By definition of $C^+_i$, immediately before iteration $i$ of
line~\ref{line:to-ai} there are no $C^+_i$-to-$\{C \setminus C^+_i\}$
residual paths. By \sources($C^+_i,C \setminus C^+_i,C^+_i$), there are
no $C^+_i$-to-$\{C \setminus C^+_i\}$
residual paths immediately after iteration $i$ of
line~\ref{line:to-ai} as well. This shows that there are no $X$-to-$\{C \setminus C^+_i\}$
residual paths at that time.

The flow pushed in line~\ref{line:from-ai} can be decomposed into a flow whose sources and
sinks are all in $C^-_i$ and a flow whose source is $a_i$ and whose
sinks are all in $C_i^-$. By \sinks($C^+_i,C \setminus C^+_i,C^-_i$),
there are no $C^+_i$-to-$\{C \setminus C^+_i\}$
residual paths immediately after iteration $i$ of
line~\ref{line:from-ai}. This shows that there are no $X$-to-$\{C \setminus C^+_i\}$
residual paths at that time. Hence $C^+_{i+1} \subseteq C^+_i$, as desired.

The proof of the analogous claim for $C^-_{i+1}$ is similar.
\end{proof}

\begin{lemma}
Just after line~\ref{line:to-ai} of Algorithm~\ref{alg:msms} is executed in iteration $i$, 
$C_i^+ \nrightarrow C_i^-$, $C_{i+1}^+
\nrightarrow \{a_j\}_{j \leq i}$, $\{a_j\}_{j<i} \nrightarrow C_i^-$.
\end{lemma}
\begin{proof}
The flow pushed in line~\ref{line:to-ai} can be decomposed into a flow whose sources and
sinks are all in $C^+_i$ and a flow whose sources are in $C_i^+$ and
whose sink is $a_i$.
\begin{itemize}
\item $C_i^+ \nrightarrow C_i^-$ by \sources($C_i^+,C_i^-,C^+_i$) 
\item $C_{i+1}^+ \nrightarrow \{a_j\}_{j < i}$ by \sources($C_{i+1}^+,a_j,C^+_i$)
\item $C_{i+1}^+ \nrightarrow \{a_i\}$ by definition of \proc{CycleToSingleSinkLimitedMaxFlow}
\item $\{a_j\}_{j<i} \nrightarrow C_i^-$ by \sources($a_j,C_i^-,C^+_i$) 
\end{itemize}
\end{proof}

\begin{lemma}
Just after line~\ref{line:from-ai} of Algorithm~\ref{alg:msms} is executed in iteration $i$, 
$C_i^+ \nrightarrow C_i^-$, $C_{i+1}^+
\nrightarrow \{a_j\}_{j \leq i}$, $\{a_j\}_{j \leq i} \nrightarrow C_{i+1}^-$.
\end{lemma}
\begin{proof}
The flow pushed in line~\ref{line:from-ai} can be decomposed into a flow whose sources and
sinks are all in $C^-_i$ and a flow whose source is $a_i$ and whose
sinks are all in $C_i^-$.
\begin{itemize}
\item $C_i^+ \nrightarrow C_i^-$ by \sinks($C_i^+,C_i^-,C^-_i$)
\item $C_{i+1}^+ \nrightarrow \{a_j\}_{j \leq i}$ by \sinks($C_{i+1}^+,a_j,C^-_i$)
\item $\{a_j\}_{j<i} \nrightarrow C_{i+1}^-$ by \sinks($a_j,C_{i+1}^-,C^-_i$) 
\item $a_i \nrightarrow C_{i+1}^-$ by definition of \proc{SingleSourceToCycleLimitedMaxFlow}
\end{itemize}
\end{proof}

Let $C^+ (C^-)$ denote the set of nodes in $C$ with positive
(negative) inflow just before line~\ref{line:pseudo} is executed.
\begin{corollary}\label{cor:before-pseudo}
Just before line~\ref{line:pseudo} of Algorithm~\ref{alg:msms} is executed, 
$C^+ \nrightarrow C^-, C^+ \nrightarrow A, A\nrightarrow C^-$.
\end{corollary}

\begin{lemma}\label{lem:final}
The following are true upon termination:
\begin{enumerate}
\item $\fvec$ is a pseudoflow
\item  $\fvec$ obeys conservation everywhere except at $S,T,A$
\item  $ S \stackrel{G_{\fvec}} \nrightarrow T, S \stackrel{G_{\fvec}}
  \nrightarrow A, A
\stackrel{G_{\fvec}} \nrightarrow T$.
\end{enumerate} 
\end{lemma}
\begin{proof}
Since every addition to $\fvec$ along the algorithm respects
capacities of all darts, 
$\fvec$ is a pseudoflow at all times.
By induction, the only nodes that do not obey conservations after the
recursive calls are those of $S,T$ and $A$. Subsequent changes to
$\fvec$ only violate conservation on the nodes of $C$, but any such
violation is eliminated in line~\ref{line:pseudo}. Therefore upon
termination $\fvec$ obeys conservation everywhere except $S,T$ and $A$.

Since, by Lemma~\ref{lem:between-recurse-and-pseudo} and Corollary~\ref{cor:before-pseudo} before
line~\ref{line:pseudo} $C^+ \nrightarrow A$
and $C \nrightarrow T$, the flow pushed back from
$C^+$ in line~\ref{line:pseudo} must be pushed back to $S$. Similarly,
the flow pushed back to $C^-$ must be pushed back from $T$.

Let $f_+$ ($f_-$) be the flow pushed back from $C^+$ to $S$ (from $T$
to $C^-$) in line~\ref{line:pseudo}.
Consider first pushing back $f_+$.
\begin{itemize}
\item $S \nrightarrow T $ by \sources($S,T,C^+$)
\item $S \nrightarrow A$ by \sources($S,A,C^+$)
\item $A \nrightarrow T$ by \sources($A,T,C^+$)
\item $S \nrightarrow C^-$ by \sources($S,C^-,C^+$)
\item $A \nrightarrow C^-$ by \sources($A,C^-,C^+$)
\end{itemize}

Next consider pushing $f_-$
\begin{itemize}
\item $S \nrightarrow T $ by \sinks($S,T,C^-$)
\item $S \nrightarrow A$ by \sinks($S,A,C^-$)
\item $A \nrightarrow T$ by \sinks($A,T,C^-$)
\end{itemize}
\end{proof}

\section{Correctness of Algorithm~\ref{alg:fix}}\label{sec:abs-cor}

\begin{lemma}\label{lem:invs}
The following holds immediately after iteration $i$ of the main loop (line~\ref{0-loop}).
\begin{enumerate}
\item For $j \leq i, j' >j$, if $p_j$ has positive inflow, there is no
  residual path from $p_j$ to $p_{j'}$.
If $p_j$ has negative inflow, there is no residual path from $p_{j'}$ to $p_j$.\label{inv:proc1}
\item For $j,j' \leq i$, if $p_j$ has positive inflow and $p_{j'}$ has negative inflow then there is no $p_j$-to-$p_{j'}$ residual path.\label{inv:proc2}
\end{enumerate}
\end{lemma}
\begin{proof}
By induction on the number of iterations $i$ of the loop. 
For $i=0$ the invariants are trivially satisfied.

First note that the adjustments to capacities and flow in lines~\ref{0-cap}--\ref{0-reduce-flow} do not create any new residual paths since capacities are only reduced, and no residual capacity increases. Therefore, it suffices to argue just about the flow pushed in line~\ref{0:flow}.

Assume the invariants hold up until the beginning of the $i^{th}$
iteration.  We show that the invariants hold at the end of the iteration.
Suppose that $p_i$ has positive inflow at the end of the $i^{th}$ iteration (the case of
negative inflow is similar).
\begin{enumerate}
\item  Since the flow pushed in line~\ref{0:flow} is
  limited by $|\inflow(p_i)|$, the fact that $p_i$ has positive inflow at the end implies that the flow pushed was a maximum flow from $p_i$ to $p_{i+1}$. 
Since the capacities of darts between $p_{k+1}$ and $p_k$ for $k>i$ are sufficiently large, 
the maximality of the flow implies that there are no $p_i$-to-$p_k$ residual paths for any $k>i$.

The invariant holds for nodes $p_j$ with positive inflow and 
  $j<i$ by \sinks($\{p_j\}$,  $\{ p_{j'} : j'>j \}$, $\{p_{i+1} \}$).

\item Invariant~\ref{inv:proc2} holds for $j,j'<i$ by \sinks(
$\{p_j : j<i, \inflow(p_j)>0\}$, $\{p_j : j<i, \inflow(p_j)<0\}$,
$\{p_{i+1}\}$).

The invariant holds for $p_i$  by invoking the \sources($\{p_i\}$, $
\{p_j : j<i, \inflow(p_j)<0\}$, $\{p_i\}$).
\end{enumerate}
\end{proof}

\begin{corollary}
Algorithm~\ref{alg:fix} is correct
\end{corollary}
\begin{proof}
The flow $\fvec'-\fvec_0$ satisfies conservation everywhere except at nodes of $P$ since the algorithm only pushes flow between nodes of $P$.
By Lemma~\ref{lem:invs}, with respect to $\fvec'$, there are no residual paths from nodes
  of $P$ with positive inflow to nodes of $P$ with negative inflow.
\end{proof}

\section{Converting a Pseudoflow into a Maximum Feasible Flow} \label{apn:psuedoflow-to-flow}

Let $\fvec$ be a pseudoflow in a planar graph $G$ with node set $V$, sources $S$ and sinks $T$. Let
$V^+$ denote the set of nodes $\{v \in V \setminus (S \cup T) :
\inflow_{\fvec}(v) > 0\}$. Similarly, let
$V^-$ denote the set of nodes $\{v \in V \setminus (S \cup T) :
\inflow_{\fvec} (v) < 0\}$.
Suppose $S \cup V^+ \stackrel{G_{\fvec}}\nrightarrow T \cup V^-$.
In this appendix we show how to convert $\fvec$ into
a maximum feasible flow $\fvec'$.
This procedure was first described for planar graphs by Johnson and
Venkatesan~\cite{JV82}. The original description of the procedure took
$O(n \log n)$ time, but using the flow cycles canceling technique of
Kaplan and Nussbaum~\cite{KaplanNussbaum2009} the running time is $O(n)$.

We begin by converting $\fvec$ to an acycic pseudoflow in linear time
\cite{KaplanNussbaum2009}. That is, after the conversion there is no
cycle $C$ such that $\fvec[d] > 0$ for every dart $d$ of $C$. Since
$\fvec$ is acyclic, the darts with a positive flow induce a
topological ordering on the nodes of the graph $G$. Let $v$ be the
last member of $V^+$ in the topological ordering, and let $d$ be an
arbitrary dart that carries flow into $v$. We set $\fvec[d] =
\max\{\fvec[d] - \inflow_{\fvec} (v), 0\}$, and set $\fvec[\rev(d)]$
accordingly. The flow assignment $\fvec$ maintains the invariant $S
\cup V^+ \stackrel{G_{\fvec}}\nrightarrow T \cup V^-$ by \sinks($S
\cup V^+$,$T \cup V^-$,$\{v\}$). As long as $v$ is in $V^+$, there
must be a dart $d$ which carries flow to $v$. By changing the flow on
$d$ we cannot add to $V^+$ a new node that appears later than $v$ in
the topological ordering. We repeat this process until $V^+$ is
empty. Since we reduce the flow on each dart at most once, this takes
linear time. Next we handle $V^-$ while keeping the invariant $S \cup
V^+ \stackrel{G_{\fvec}}\nrightarrow T \cup V^-$ in a symmetric way,
by repeatedly fixing the first vertex if $V^-$ in the topological
ordering. 

The total running time is $O(n)$, and since $V^+, V^-$ are both empty,
we get from the invariant $S \cup V^+ \stackrel{G_{\fvec}}\nrightarrow
T \cup V^-$ that the resulting pseudoflow is a feasible flow. This is
the required flow $\fvec'$.

\newpage
\section{Pseudocode for \proc{FixConservationOnPath}}\label{sec:fix-code}

\begin{algorithm}[h!]
\caption{\proc{FixConservationOnPath}$(G,P, \cvec, \fvec_0)$} \label{alg:efficient-fix-full}
{\bf Input:} a directed planar graph $G$, simple path
$P=d_1 d_2\ldots d_k$, capacity function $\cvec$, and a pseudoflow $\fvec_0$\\
{\bf Output:} a pseudoflow $\fvec$ s.t. (1)
$\fvec-\fvec_0$ satisfies conservation at nodes not on $P$ and (2)
w.r.t $\fvec$, no residual path exists from $\{ v \in P :
\inflow(v)>0\}$ to $\{ v \in P : \inflow(v)<0\}$. \\
\begin{algorithmic}[1]
\STATE \vspace{-10pt}let $X^*$ be the set of endpoints in the planar dual $G^*$ of the darts of $P$
\STATE initialize to zero an array $\phivec^X[\cdot]$ indexed by the
nodes of $X^*$ 
\STATE let $H^*$ be the graph obtained from $G^*$ by removing the darts of $P$
\STATE compute, using the MSSP algorithm~\cite{Klein05,CabelloC07}, a table
$\DIST^*[\cdot,\cdot]$ of distances in $H^*$ between nodes of $X^*$
w.r.t the lengths $\cvec-\fvec_0$ \label{FCOP-MSSP}
\STATE $\fvec := \fvec_0$\label{FCOP-f0}
\STATE  $\cvec[d] := \cvec[d] + M$ for all darts of $P\cup \rev(P)$
\STATE initialize an array of length $k$ by $\vvec[i] := $ inflow at $p_i$ with respect to $\fvec$

\FOR {$i=1,2, \ldots, k$}\label{FCOP-mainloop}
   \STATE let $p_i$ and $p_{i+1}$ be the tail and head of $d_i$,
   respectively
   \COMMENT{ \vspace{10pt} \\ \small \% reduce
    the capacities of $d$ and $\rev(d)$ by $M$ and adjust the flow appropriately}
   \FOR{$d \in \set{ d_i, \rev(d_i)}$} \label{1-loop-drevd}
     \STATE $\cvec[d] := \cvec[d] - M$ 
      \STATE $\text{\it old flow} := \fvec[d] + \phivec^X[\head_{G^*}(d)] - \phivec^X[\tail_{G^*}(d)]$ \label{FCOP-old-flow}
      \STATE $\text{\it new flow} := \min \set{\text{\it old flow}, \cvec[d]}$ \COMMENT{ adjusted flow must not exceed adjusted capacity}
      \STATE $\fvec[d] := \text{\it new flow} - \phivec^X[\head_{G^*}(d)] +
   \phivec^X[\tail_{G^*}(d)]$ \COMMENT{the explicit flow does not include
     the circulation component}
      \STATE $\fvec[\rev(d)] := - \fvec[d]$\label{FCOP-reduce-flow}
      \STATE $\vvec[\head(d)] := \vvec[\head(d)] + \text{\it new flow} - \text{\it old flow}$ \COMMENT{ update the inflow at $p_i$ and $p_{i+1}$}
      \STATE $\vvec[\tail(d)] := \vvec[\tail(d)] - \text{\it new flow} + \text{\it old flow}$
\COMMENT{\vspace{6pt}}
   \ENDFOR
   \STATE \algorithmicif { $ \vvec[p_i] > 0$} \algorithmicthen \ $d
   := d_i$
   \algorithmicelse \ $d := \rev(d_i)$ \label{FCOP-test-inflow}
   \COMMENT{find in which direction flow should be pushed
     \vspace{10pt}\\\hspace{-2pt} \small \% push flow on $d$ to make
     its residual capacity zero as required for Hassin's algorithm}
   \STATE {\it residual capacity} $:= \cvec[d] - \fvec[d] -\phivec^X[\head_{G^*}(d)] + \phivec^X[\tail_{G^*}(d)]$\label{FCOP-residual-capacity}
   \STATE {\it val} $:= \min \set{\text{\it residual capacity}, |\vvec[p_i]|}$ \COMMENT{ amount of flow pushed on $d$}
   \STATE $\fvec[d] = \fvec[d] +  \text{\it val}$ ; $\fvec[\rev(d)]
   := -\fvec[d]$ 
\label{FCOP-push-on-revd}
   \STATE $\vvec[\tail(d)] = \vvec[\tail(d)] - \text{\it val}$ ; $\vvec[\head(d)]  = \vvec[\head(d)] +   \text{\it val}$ \COMMENT{ update the inflow at $p_i$ and $p_{i+1}$
\vspace{6pt}\\ \small \% push excess inflow from $\tail(d)$ to $\head(d)$
     using Hassin's algorithm}
   \STATE let $\ell[d'] := \cvec[d'] - \fvec[d']$ for all darts $d'
   \in P\cup \rev(P)$ \label{FCOP-residual-capacities} \COMMENT{\parbox{2.4in}{lengths
     of explicit darts are residual
     capacities (not including circulation component) }} \vspace{2pt}
   \STATE $\ell[\rev(d)] := |\vvec[p_i]|
   -\phivec^X[\tail_{G^*}(\rev(d))] + \phivec^X[\head_{G^*}(\rev(d))]$
   \COMMENT{ \parbox{2.2in}{set the limit on the residual capacity of $\rev(d)$
     (adjusted by circulation component)}} \vspace{2pt}
   \STATE $\phivec_i^X(\cdot) := FR(\DIST^* , \ell, \phivec^X,
   \head_{G^*}(d))$ \label{FCOP-FR}
\COMMENT{\parbox{3in}{face potential are distances in $G^*$ from $\head_{G^*}(d)$
  w.r.t.\  the reduced lengths induced by $\phivec^X$} }
   \STATE {\it val} $:= \phivec_i^X[\head(\rev(d))] - \phivec_i^X[\tail(\rev(d))]$\label{FCOP-val} \COMMENT{the amount of flow
     assigned to $d$ by the circulation  corresponding to $\phivec_i^X$}
   \STATE $\fvec[d] := \fvec[d] + \text{\it val}$ ; $\fvec[\rev(d)] :=
   -\fvec[(d)]$\label{FCOP-not-on-revd2} \COMMENT{ do not push the
     circulation on $d$ and $\rev(d)$}
   \STATE $\phivec^X  = \phivec^X + \phivec_i^X$ \COMMENT{ accumulate the current circulation}
   \STATE $\vvec[\tail(d)] = \vvec[\tail(d)] - \text{\it val}$ ; $\vvec[\head(d)]  = \vvec[\head(d)] +   \text{\it val}$ \COMMENT{ update the inflow at $p_i$ and $p_{i+1}$
\vspace{6pt}\\ \hspace{-10pt}\small \%
     find a feasible circulation}
\ENDFOR
\STATE let $x$ be an arbitrary node in $G^*$
\STATE let $\ell[d] := \cvec[d] - \fvec[d]$ for all darts $d \in G$ \label{FCOP-lastell}
\STATE $\chivec := \proc{SingleSourceShortestPaths}(G^*, \ell, x)$\label{FCOP-SSSP} \COMMENT{ distances of the nodes of $G^*$ from $x$}
\STATE $\fvec'[d] := \fvec[d] + \chivec[\head_{G^*}(d)] - \chivec[\tail_{G^*}(d)]$ for every dart $d \in G$ \COMMENT{add explicit flow and implicit circulation}
\RETURN $\fvec'$
\end{algorithmic}
\end{algorithm}

\section{Alternative to Line~\ref{FCOP-SSSP} of Algorithm~\ref{alg:efficient-fix-full}} \label{sec:alternative}
In this section we show that it is not necessary to use a generic
shortest path algorithm that works with negative lengths to compute
a feasible circulation $\chivec$ in line~\ref{FCOP-SSSP} of Algorithm~\ref{alg:efficient-fix-full}.
Instead of choosing $x$ to be an arbitrary node in $G^*$, let $x$ be
an arbitrary node of $X^*$. Let $\chivec(y)$ denote
the $x$-to-$y$ distance in $G^*$ w.r.t.\ the lengths $\ell$.
Instead of computing $\chivec$ using a shortest path algorithm that accepts
negative lengths, we will compute it more efficiently in two
steps. Pseudocode is given below as Algorithm~\ref{alg:alternative}.
 In the first step we use FR to compute the distances to just the
nodes of $X^*$. In the second step we extend these distances to all
other nodes using Dijkstra's algorithm.

In the first step the algorithm computes distances in $G^*$ from $x$
to all nodes of $X^*$ w.r.t.\ $\ell_{\phivec^X}$, the reduced
lengths of $\ell$ induced by the feasible price function $\phivec^X$.
This is done by an additional invocation of FR (line~\ref{FCOP-lastFR}). 
Let $\psivec^X$ denote these distance labels.
By definition of reduced lengths and a telescoping sum similar to the
one in the derivation of
Eq.~\eqref{eq:Q-red-len}, 
\begin{equation}
\psivec^X[y] = \chivec(y) + \phivec^X[x] - \phivec^X[y].
\end{equation}
Since both $\phivec^X$ and $\psivec^X$ are known, the algorithm can
compute the unreduced distances $\chivec[y]$ for all $y \in X^*$
(line~\ref{FCOP-convert}).
Next, the algorithm runs Dijkstra's algorithm on $G^*$, initializing
the label of each node $y$ of $X^*$ to its correct value
$\chivec[y]$. Since in the dual the darts of $P \cup \rev(P)$ are only incident
to nodes of $X^*$, Dijkstra's algorithm initialized in this manner
correctly outputs the 
distance labels for all nodes of $G^*$ even if some of the darts
of $P \cup \rev(P)$ may have negative lengths.

\begin{algorithm}\caption{ Alternative to line~\ref{FCOP-SSSP} of Algorithm~\ref{alg:efficient-fix-full} \label{alg:alternative}}
\begin{algorithmic}[1]
\STATE let $x$ be an arbitrary node in $X^*$
\STATE $\psivec^X = FR(\DIST^* , \ell, \phivec^X, x)$ \label{FCOP-lastFR} \COMMENT{find distances of the nodes of $X^*$ from $x$ w.r.t.\ price function $\phivec^X$}
\STATE initialize to $\infty$ an array $\chivec$ indexed by the nodes of $G^*$
\STATE $\chivec[y] = \psivec^X[y] -\phivec^X[x] + \phivec^X[y]$ for every $y \in X^*$ \label{FCOP-convert}\COMMENT{ distances of the nodes of $X^*$ from $x$}
\STATE extend $\ell$ to all darts of $G^*$ by $\ell[d] :=\cvec[d]-\fvec_0[d]$ for every dart $d \in G$
\STATE $\chivec := \proc{Dijkstra}(G^*, \ell, x, \chivec)$\label{FCOP-dijkstra} \COMMENT{ distances of the nodes of $G^*$ from $x$}
\end{algorithmic}
\end{algorithm} 

\section{Correctness of Algorithm~\ref{alg:lim_cyc}}\label{apn:c2a-cor}
\begin{lemma}
Algorithm~\ref{alg:lim_cyc} is correct.
\end{lemma}
\begin{proof}
Any flow that is pushed by the algorithm originates at $C^+$ and
terminates at $C^+ \cup \{t\}$. Therefore, $\fvec - \fvec_0$ violates
conservation only at $C^+ \cup \{t\}$. 
In line~\ref{CTS:pushback}, flow of $\fvec$ is pushed back so that no node of
$C^+$ has negative inflow w.r.t.\ $\fvec_0 + \fvec$.
It is possible to do so by only pushing back flow of $\fvec$ (rather
than flow of $\fvec_0 + \fvec$) since by assumption no node of $C^+$ has negative
inflow w.r.t.\ $\fvec_0$.

By maximality of the flow pushed in line~\ref{CTS:stmaxflow},
just after  line~\ref{CTS:stmaxflow} is executed there are no
$C^+$-to-$t$ residual paths. Clearly, this remains true when the capacities of
the artificial darts are set to zero. 
In line~\ref{CTS:fix} flow is pushed among the nodes of $C^+$, so by
\sources($C^+,t,C^+$), there are no $C^+$-to-$t$ residual paths after
line~\ref{CTS:fix} either. Moreover, by definition of
\proc{FixConservationOnPath}, there are no $C_{>0}$-to-$C_{<0}$ residual
paths immediately after line~\ref{CTS:fix} is executed, where $C_{>0}$
($C_{<0}$) is the set of nodes of $C^+$ with positive (negative)
inflow at that time. 
Line~\ref{CTS:pushback} pushes flow into $C_{<0}$, making all the
nodes of $C_{<0}$ obey conservation. By \sinks($C_{>0},t,C_{<0}$)
there are no $C_{>0}$-to-$t$ residual paths upon termination of the
procedure.
This completes the proof since $C_{>0}$ is the set of nodes of $C$
with positive inflow upon termination.
\end{proof}

\end{appendix}
\end{document}